# Modeling Multi-Point Transport Protocol in P2P Networks

## Yong-Xiang Zhao[*], Chang-Jia Chen

School of Electronic and Information, Beijing Jiaotong University, Beijing 100044, China

**Abstract:** Traditional end-to-end congestion control mechanisms assume data transferring happens between each pair user. In contrast, in a P2P network, many peers may locally keep a copy of a specific data object. If the path between a pair of peers is congested, the requesting peer who wants to download data will switch to another peer in its neighbor peer list to fetch the data instead of decreasing the download rate from the current peer. Thus, it is critical to study the performance in multi-point-to-multi-point (M2M) transport protocol in a P2P network. In this paper, we build a mathematical model for identifying the key parameters for the M2M transport protocol and also the relationships among these parameters. Finally, we conduct simulation experiments to validate our model.

## 1. Introduction

End-to-end congestion control mechanism plays a key role for the stability of the Internet [1]. The main objective of a congestion control mechanism is to dynamically adjust the source to send data at a rate according to the available bandwidth on the path between the source and destination node. The assumption behind traditional congestion control mechanisms is that only one node has the data that the requesting node wants. So the data transfer happens only between a pair of nodes. We name this kind of mechanism as end-to-end congestion control (E2E-CC). The basic function in E2E-CC is to determine how frequently an end node holding the data should send out packets to the other end node who wants the data.

In most popular P2P applications, a specific file is in general stored at multiple peer nodes. A user can download the data of this file from some of these peers concurrently. For example, Bittorrent downloads a file from four peer nodes in parallel with four TCP connections. Both PPstream and PPlive also use this kind of method to download data. Every peer in a P2P streaming system (e.g., PPstream or PPlive) is connected to tens or even hundreds of peers simultaneously.

The common communication pattern among these P2P applications is to swarm multiple peers at the application layer to distribute a data file or a live video

This work was supported by the National Natural Science Foundation of China (Nos. 60672069, 60772043) and China 973 2007CB307101.
*Corresponding author Address: School of Electronic and information, Beijing jiaotong University, Beijing 100044, China. Tel: +86 10 51684759. Fax: +86 10 51683682
E-mail address: yxzhao2009@gmail.com (Yong-Xiang. Zhao).

collectively. The key for highly effective data distribution in a P2P environment is due to the fact that, at any time, every node holds certain data that is desired by other nodes, and each node can fetch data from many different nodes simultaneously. Thus, a common transport problem among these P2P applications is to determine to download which pieces of data from which neighboring peers concurrently and how frequently. We call this transport problem as multi-nodes-to-multi-nodes (M2M) transport problem, which is also called multi-point-to-multi-point transport problem. Almost all the existing work in the area of P2P volume data dissipations only consider the aspects of "which and who" strategies [3,4] that determine to download which piece of data from which neighbor peers in terms of data fetching. They consider the availability of contents as the key to achieving efficient transport. However, how often to download the data is also an important problem in P2P volume data dissipations. This has not been realized before. Just like in the case of E2E-CC, there is a congestion control problem in M2M transport. In this paper, we assume the availability of contents is not a problem and concentrate primarily on the congestion control problem in M2M transport. We model the M2M transport problem analytically and find that, given network conditions and quality of service requirements in M2M transport, there is a maximum throughput.

M2M transport can be seen to work on top of many E2E connections (either TCP or UDP connections). We name the protocol used by each of those E2E connections as the underlying protocol on which the M2M transport works. UDP seems to be a more appropriate underneath protocol than TCP since almost all P2P streaming networks (e.g., live and VoD) have migrated the underneath protocol to UDP in their M2M transport protocol based on our measurements [10]. Moreover, BitTorrent is currently trying to do similar actions [5]. Ref. [6] claims "TCP is still dominating in terms of packet and byte numbers, while UDP is responsible for the main part of traffic flows. A port-based analysis suggests that the recent increase in UDP flows stems mainly from P2P applications using UDP for their overlay signaling traffic."

Facing such a large scale UDP "abuse", the same worry as in [2] has been raised again by many researchers. The fact that the Internet does not collapse after such "abuse" may suggest the following two important points. First, the network congestion mechanism in M2M transport is different from that in E2E transport. We will explain this point in the following paragraph. Second, certain congestion control can be implemented within the M2M transport protocol naturally. For example, in E2E-CC, an end node does not control the path selection in general. Thus, the transport path cannot be changed if one of its links on the path is congested. However, in a P2P application, if a link between a pair of nodes is congested, the requesting node could switch to another node (peer) to download the requested data smoothly. We believe that such change of requested peers according to end-to-end transfer quality is also a kind of congestion control, which is implemented on a per-connection basis instead of on a per-packet basis. M2M transport protocol can also limit each peer's traffic-generating rate using a mechanism such as slide windows.

Figure 1 illustrates where congestion can happen in the M2M architecture. Request user wants to download a file with a size of S. All user-1, user-2, …, and user-$2^n$ have this file. We assume the underlying network uses the shortest path routing algorithm to determine the route between each pair of users. From the viewpoint of the Request user, it is connected to user-$i$ ($1 \leq i \leq 2^n$) via a tree generated by using the shortest path algorithm. In E2E-CC transport protocol, to download such a file, every link on the path from the Request user to the destination has to transfer S bytes. If two flows running E2E-CC pass through the same link, E2E-CC has to throttle the sending rate in order to let the two flows share the link fairly. In M2M transport protocol, the Request user can download the file from the $2^n$ users in parallel. The Request user can divide the file into $2^n$ pieces, and download them from the $2^n$ users in pipeline. From Figure 1, it is seen that the access link of the Request user has to transfer a total amount of S bytes. Each link in Stage 1 only needs to transfer half of the file. Each link in Stage 2 only needs to transfer 25 percent of the file, and so on. The farther a link is away from the Request user, the less the traffic that the link needs to carry. In short, the M2M transport protocol can distribute traffic uniformly on the network. The closer to a peer, the more traffic is expected to be exerted onto the network. For a particular user, its access link carries the heaviest traffic load. On the other side, most residential users access the Internet via links with limited bandwidth. This means the access rate of a user is much smaller than the capacity of a link in the backbone network. Since the access link has to transport all the traffic to and from a user and this link has much smaller bandwidth than a link in the backbone network, it is reasonable to assume that access link would become the bottleneck link in the M2M transport protocol. Hence, in this paper, we assume the congestion will only happen at the access links of peers and links on the backbone network are congestion free.

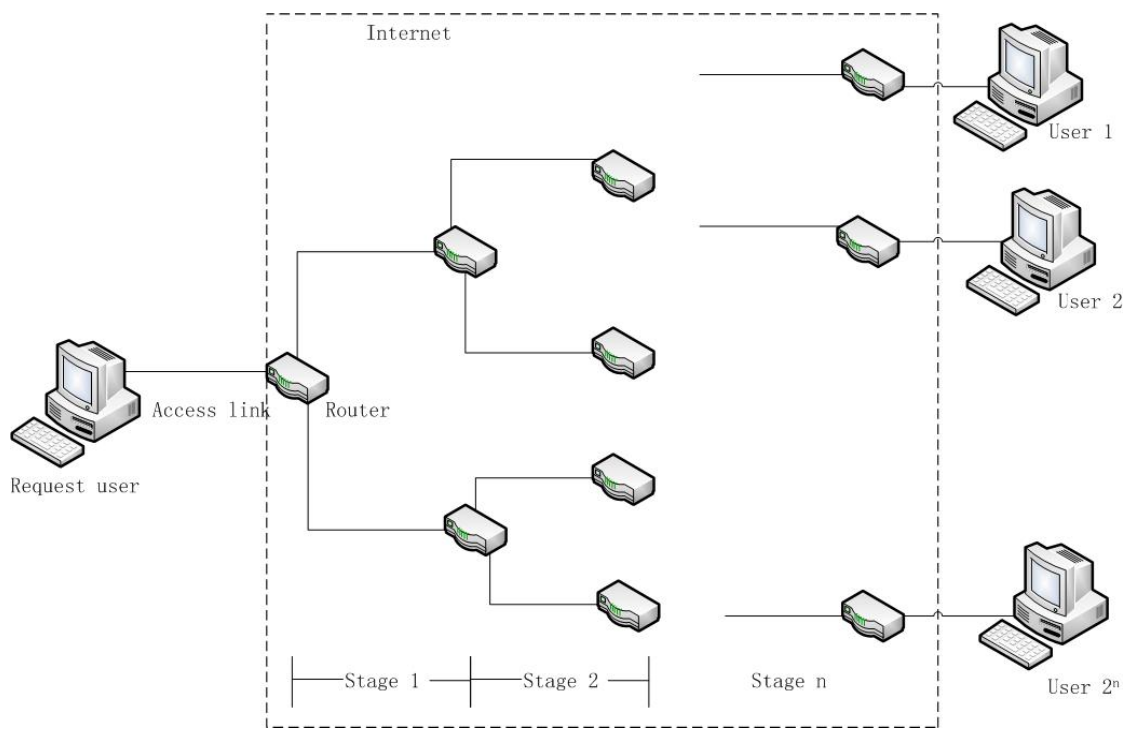

Figure 1: Tree generated by the shortest path algorithm.

We discuss the differences between M2M transport protocol and traditional congestion control mechanism. Firstly, M2M transport protocol can spread traffic to overall Internet which can relieve congestion in the backbone networks greatly. In the example shown in Figure 1, the traffic passing through a specific link decreases exponentially with the min-hop distance from the Request user to the specific link. Although the traffic from two Request users may pass across the same link in the backbone network, the sum of two such user's traffic is relatively small compared to the capacity of a link on the backbone network. So the interference between the two request users will be very small and thus negligible. Secondly, M2M transport protocol can run away from a congested link by switching peer to be communicated. If a link in the backbone network is congested, a request user who happens to fetch traffic through this link can switch to another peer to avoid the use of this link. In the traditional TCP, each link on the path between a pair of TCP end users has to carry all the data traffic wanted by the requesting user. If two paths carrying TCP flows pass through the same link, they have to adjust their rate to share the link fairly. That means tradition TCP has to spread traffic over time when one or more links on the path used by the TCP flow is congested. In short, we think the mechanism of M2M has some kinds of congestion control function. Even eventually M2M cannot avoid the congestion collapse of the Internet, it can at least delay the coming of the collapse.

MTCP[9] assumes there are multiple paths between a pair of node and tries to balance the traffic among these path. However, if the number of paths is large enough, the access link will become the bottleneck link. Ref. [11] tries to solve the "TCP collapse" problem in data center. This problem happens in the following situation: when multiple nodes transfer data to a source node simultaneously, the access link of the source node will be congested. In this paper, the same line is followed. We assume a node can download a data file from a large number of nodes in parallel. Since the traffic is spread on the whole network, we assume the congestion happens in the access network instead in the backbone network. Furthermore, as found in [10], PPStream has migrated its underlying transport protocol to UDP in 2008. Thus, we assume that a P2P application adopts UDP as underlying its transport protocol and discuss the congestion problem when peers download data from each other in P2P environment.

Since many P2P applications tend to use UDP as their transport protocol to download data. A natural question is how often a peer should download data from other peers in the context of UDP traffic. Ref. [1] points out that the optimal number of peer neighbors is between 15 and 32 without explanation. The model in this paper will explain this phenomenon to certain extents. In this paper, we model the relationship among network settings, packet delay, and download rate. We find there exists an optimal download rate for given network settings and delay requirement. This means we should adjust the download rate to the optimal value according to network settings and delay requirement when designing a real M2M transport protocol.

The rest of this paper is organized as follows. In Section 2, we introduce M2M transport mechanism and then build our model. In Section 3, we conduct simulation experiments to valid the built model. In Section 4, we conclude this paper.

**2. An Analytical Model for M2M Transport**

*2.1 A simple M2M transport protocol*
To concentrate on the M2M transport issue in P2P volume data dissipation, we ignore the data availability problem deliberately. This means each node always has the data that other nodes need. We assume there are totally N nodes in a P2P system. All these nodes are connected to the Internet via access links with limited bandwidth. Every node behaves as a client to download data from other nodes and also behaves as a server to upload data to other nodes. The request node will send a request to the destination node, and then the destination node will reply a data packet to the request node. Each node wishes to download data as quickly as possible.

At each node, there are M running threads in M2M transport. Each thread runs a protocol similar to the traditional stop-wait protocol. This means after the thread sends out a request, it has to wait until one of the following events happens: Either the arrival of a requested data packet or a timeout event. Then the thread can send a new request. The destination of a request is randomly selected from the N peer nodes. A peer can have at most M pending requests at any time. Then the peer has to stop until any wanted data packet comes in or timeout event happens. So the number of parallel threads, M, is also known as the window value in this case. Note that if a data packet arrives at the request node after its corresponding request is timeout, this data packet is simply dropped. The reason behind this assumption is as follows: In a stream application such as PPlive or PPstream, a data is useless if it arrives late.

Upon receipt of a request, a node will always respond with a data packet to the initiating node of the request. We assume the size of a request packet is much smaller than that of a data packet.

The differences between the above M2M protocol and the traditional stop-wait protocol lie in the following two respects: First, the destination of each request is randomly selected; Second, we assume a thread that has just experienced a timeout event will issue a brand-new request instead of re-generating a new copy of the original request as done in a traditional stop-wait ARQ protocol. This means M2M transport protocol only cares about the data rate at which a peer receives data rather than the data reliability.

We should point out here that many P2P applications have already used multiple UDPs as their transport protocol. We use multiple threads of stop-wait protocol to describe their common communication pattern. Strictly speaking, the mechanism behind the pseudo codes presented in this paper may have already been used by many P2P applications. The mechanism itself is trivial. The contribution of this paper is to model the mechanism and find the relationship among the parameters in the model. According to the result derived using this model, we can explain the problem raised in Ref. [1]. Since the model and simulation are based on this M2M protocol, we describe the protocol in detail in following paragraph.

Figure 2 gives the pseudo codes of the above model. Besides the common IP packet header, we need to insert a M2M packet header in the payload of each packet. The M2M packet header includes two fields: request_data_ID field and request_ID field. In Figure 2, Lines 1-4 shows the structure of M2M packet header. The request_data_ID field represents which piece of data this request packet wants. The request node has a global variable Local_request_ID to record the largest sequence number that it has assigned for a request. Line 5 shows the definition of Local_request_ID. After sending out a request, the value of this variable will increase one. This variable's value is copied into the request_ID field of each newly generated M2M packet to uniquely identify a request packet composed by the request node. When the destination node receives this request, it will generate a corresponding data

```
1:     Struct M2M_packet {
2:         request_data_ID
3:         request_ID
4:     }
5:     External variable: Local_request_ID

6:   Thread_agent (request_data_ID) {
7:         M2M_packet request_pkt
8:         Des = select an IP randomly from the destination IP set
9:         Assign Des as the destination address of request_pkt packet
10:        request_pkt. request_data_ID = Get a request_data_ID from upper layer;
11:        Local_request_ID++;
12:        request_pkt. request_ID = Local_request_ID;
13:        send (request_pkt) and start a timer for this request;
14:        wait (event); //for a pending thread
15:        If (event== receipt of a data packet) then
                data_packet = data packet Received
16:            if (data_packet.request_ID equals to local_request_ID) then
17:                Submit data to the upper layer
18:            else
19:                drop data_packet
20:                goto 14    //continue waiting
21:            end if
22:        end if
23       goto 7      //start another request
24:    }           // end thread
```

Figure 2 Pseudo codes for M2M transport.

packet according to the request_data_ID and copy the request_ID field of the request packet into the request_ID field of corresponding data packet.

There are M threads running at each node. The codes of each thread are shown from line 6 to 25 in Figure 2. A thread composes a request packet first (see line 7). We name this packet as request_pkt. Lines 8 and 9 are to select a peer randomly and assign this peer as the destination of request_pkt. Line 10 is to get a request_data_ID from the upper layer and assigns it to the request_data_ID field of the request packet. Line 11 is to increase the value of variable Local_request_ID. Then the request_ID field of the request packet is initialized with local variable Local_request_ID in line 12. Next, the thread will send out the request packet and start a timer (see line 13). In line 14, the thread will suspend itself and wait for an event to happen: Either receipt of a data packet or time out.

Upon receipt of a data packet, the request_ID field of the data packet is compared against the Local_request_ID, which is shown in line 16. If they match, the received data packet is submitted to the upper layer application; otherwise, the received data packet is dropped and the thread goes to line 14 to continue waiting for its corresponding data packet. The lines from 16 to 21 describe this operation. The reason to add these lines is as follows: If a thread sends out a request and does not receive the corresponding data packet before its associated timer expires, then the thread sends a new request out. However, the former corresponding data packet may still come in. We treat such a packet as an invalid packet.

Either receipt of a correct data packet in time or time out, a thread will be released and go to line 7 to start sending a new request. Line 23 shows this operation.

*2.2 Performance of M2M transport protocol without timeout mechanism*

Figure 3 gives a generic network scenario. There are N users connected to the backbone network via access links. Each user runs M threads, each of which executes the pseudo codes in Figure 2. In this section, we set timeout value of M2M transport protocol to infinite. This means after a thread sends out a request, it will receive a responding data packet sooner or later. We name the access link from a node to the network as uplink and the link from the network to the node as downlink. We assume the bandwidth of uplink and downlinks are equal. C represents the bandwidth of uplink and downlink.

We define the time interval from the instant when a request packet is sent to the instant when the corresponding data packet comes back as rtt (round trip time). We assume the application always has requests needed to be sent. So each thread in a node can send 1/rtt request in every second. A node has M threads. So a node can send M/rtt requests per second. Since every node serves as both client and server, every node will receive M/rtt requests per second on average. Because each request packet will generate a data packet, each node will download and upload, on average, M/rtt data packets per second.

Now we consider how to calculate rtt. As shown in Figure 3, we assume user A sends a request to user B and user B responds a data packet to user A. rtt consists of the queue delay on the uplink at user A side, the queue delay on the downlink at user B side, the queue delay on the uplink at user B side, the queue delay on the downlink at user A side, and the Internet round trip delay (including propagation delay on the access link). We denote them as $D_{a\_up}$, $D_{b\_down}$, $D_{b\_up}$, $D_{a\_down}$, $T_P$, respectively. We assume $T_P$ is a fixed value. Then

$$rtt = T_P + D_{a\_up} + D_{b\_down} + D_{b\_up} + D_{a\_down} \tag{1}$$

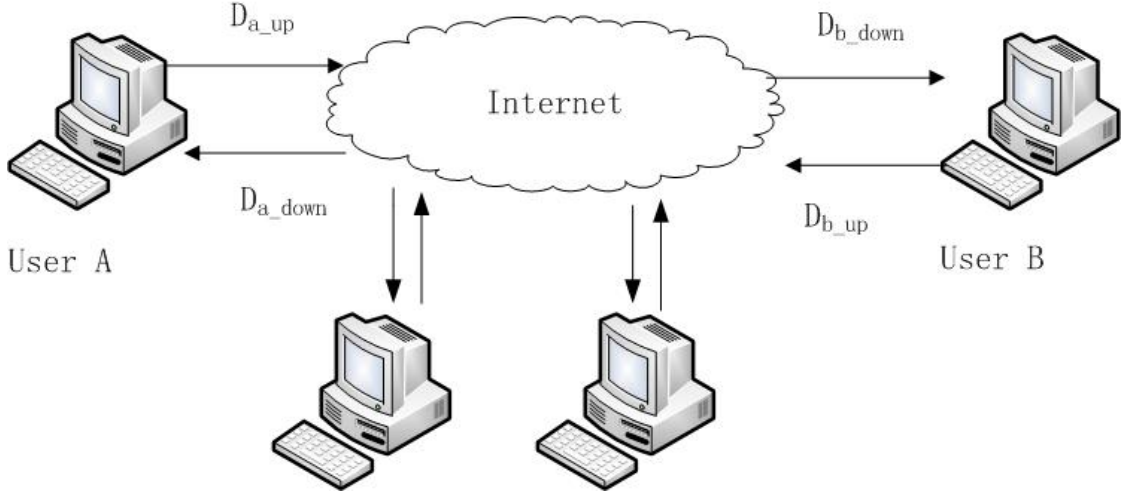

Figure 3 Delay components.

Now we discuss about $D_{a\_up}$ which is the delay that a request packet experiences on the uplink at user A side. There are two types of packets that are transmitted on this uplink: request packets sent by user A to other users in order to request data packets from them and data packets uploaded by user A as responses to other requesting users. Both of their packet-generating rates are M/rtt. Because we assume the size of a request packet is much smaller than that of a data packet, we ignore the traffic caused by request packets when we calculate delay $D_{a\_up}$. However, a request packet has to experience the same delay just as a data packet.

Let L represent the size of a data packet. Then the uplink can transmit C/L data packets per second. Let μ= C/L. we assume the transmission time of a data packet follows exponential distribution with parameter μ. Furthermore, we assume the data packet arrivals at uplink follow Poisson process. As discussed above, the packet arrival rate is M/rtt. The uplink can be modeled as an M/M/1 queue system. So $D_{a\_up}$ follows exponential distribution with parameter μ-M/rtt. Its expectation is 1/(μ-M/rtt).

For similar reasons, all $D_{b\_down}$, $D_{b\_up}$, $D_{a\_down}$ follow exponential distribution with parameter μ-M/rtt. It is reasonable to assume they are mutually independent. From (1), we have

$$E(rtt) = E(T_P) + E(D_{a\_up}) + E(D_{b\_down}) + E(D_{b\_up}) + E(D_{a\_down})$$
$$E(rtt) = T_P + \frac{4}{\mu - M/E(rtt)} \tag{2}$$

Given *μ, Tp, M*, we can get *E(rtt)* by solving this equation. This equation has two solutions. Only the solution larger than $T_P$ is valid.

The normalized throughput can be written as follows.

$$gamma = \frac{M/rtt}{\mu} \tag{3}$$

*2.3 The performance of M2M transport protocol with timeout mechanism*

In this section, we discuss M2M transport protocol with timeout mechanism. This means after a thread sends out a request, either it will receive a corresponding data packet in time or its timer finally expires. Denote P as the probability that the thread timer is finally expired and $T_{out}$ as the duration of the timer.

As discussion above, since all of $D_{a\_up}$, $D_{b\_down}$, $D_{b\_up}$, and $D_{a\_down}$ follow exponential distribution with parameter µ-M/rtt and are mutually independent. Denote the expectation of $D_{a\_up}$, $D_{b\_down}$, $D_{b\_up}$, $D_{a\_down}$ as *x*. Given the round trip time rtt and fixed Internet propagation time $T_P$, we can get *x*, from Eq. (2). That is

$$x = (rtt - T_p)/4 \tag{4}$$

Furthermore, the sum of $D_{a\_up}$, $D_{b\_down}$, $D_{b\_up}$, *and* $D_{a\_down}$ follows gamma distribution with parameters (4, *x*). Note that $T_p$ has a fixed value. If the sum of $D_{a\_up}$, $D_{b\_down}$, $D_{b\_up}$, and $D_{a\_down}$ is larger than $T_{out}$-$T_P$, the thread that sent out the request will experience a timeout. So the timeout probability P can be computed using the following equation.

$$P = 1 - gamcdf(T_{out} - T_P, 4, x) \tag{5}$$

Where gamcdf is gamma cumulative distribution function with parameters 4 and *x* at the values $T_{out}$-$T_P$.

If a thread receives a data packet after its timer is expired, this packet is invalid and will be dropped. But before being dropped, this kind of packets has already consumed certain bandwidth on the incoming access link. We use the method in [7] to model the impact caused by timeout packets. We assume *C*P* bandwidth on the access link is wasted by such invalid data packets. The valid bandwidth is thus *C-C*P*. There are *M/rtt* packet arrivals on access link. Among these packets, *M/rtt*(1-P)* packets can arrive on time. Furthermore, we assume the valid bandwidth on the access link will be used to transmit these packets. So we have

$$rtt = T_P + \frac{4}{u*(1-P) - M/rtt*(1-P)} \tag{6}$$

Given *M, $T_P$, µ*, solve Eqs. (4), (5), (6), we can get rtt, P, and x. We use the following method to solve the above equations. First, we assume rtt to be a specific value, for example, rtt= $T_P$+0.001. Step 1: From Eq. (4), we get x. Step 2: we get P with x from Eq. (5). Step 3: we get a new rtt value from Eq. (6). Step 4: If the error between the assumed value of rtt and the value gotten from Eq. (6) is larger than 0.001, we increase the assumed value of rtt with step 0.001 and repeat Steps 1-4.

The normalized throughput can be obtained from Eq. (3).

*2.4 Extend our model to the case of ADSL (asymmetric) links*

In Section 2.3, we assume users access the backbone network with symmetric access links. That is, the traffic on the uplink and that on the downlink are assumed to be equal. However, many residual users access the Internet via *ADSL* links. In this appendix, we discuss how our model works with asymmetric *ADSL* access links on which the available bandwidth on downlink is much larger than that on uplink. In this case, the uplink will become the bottleneck link. As a result, we only need to consider $D_{a\_up}$, $D_{b\_up}$, and $T_P$ in our analysis. That is

$$rtt = T_P + D_{a\_up} + D_{b\_up} \tag{7}$$

So

$$E(rtt) = E(T_P) + E(D_{a\_up}) + E(D_{b\_up})$$
$$E(rtt) = T_P + \frac{2}{\mu - M/E(rtt)} \tag{8}$$

Just as Section 2.2, Denote the expectation of $D_{a\_up}$, $D_{b\_up}$, as $x$. Solving above equation, we get *x*.

$$x = (rtt - T_p)/2$$

The sum of $D_{a\_up}$, $D_{b\_up}$ follows gamma distribution with parameters $(2, x)$. The timeout probability P can be computed using the following equation.

$$P = 1 - gamcdf(T_{out} - T_P, 2, x) \tag{9}$$

Finally, Eq. (6) will change to the following form.

$$rtt = T_P + \frac{2}{u*(1-P) - M/rtt*(1-P)} \tag{10}$$

The solving of the above equations is same as that in Section 2.3.

## 3. Analytical and Simulation Results

In this section, we first derive the analytical result of the above model. We then use simulations to validate our model. We use the following network settings. Bandwidth of access link is 512 kbps. Size of data packet follows Exponential distribution with mean 1000 bytes. μ (bandwidth of uplink) is 512k/(1000*8)=64. Tp=600ms. Timer duration is set to be 4, 3, 2 seconds. Figure 3 shows the timeout probability and normalized throughput under different Ms.

We implemented the pseudo-codes in Figure 2 in ns-2 [8]. We implemented a stop-wait thread in an agent and attach M such agents to a node to simulate the behavior of a user node. The simulation topology is shown in Figure 3, where we used a router to replace the Internet. The network settings are same as those used in the above paragraph. There are in total 110 users. Simulation time is 100 seconds. We varied the value of M to get different simulation results.

Figure 4 shows the simulation result, where x-axis is the value of M, Y-axis is normalized throughput or the probability of Timeout. Each point in the figure represents a simulation result or theoretical result. Fixing a timeout value, each curve represents simulation/theoretical result with varying M. We used three element to label a curve, i.e., (Tout=2/3/4, Sim/Theory, P/Th). The first element Tout represents the duration of Timer, which may be 2, 3, and 4 seconds, respectively. The second element Sim/Theory represents whether a curve is gotten by using simulation or analytical model. The third element P/Th represents whether a curve is timeout probability or throughput, where P stands for probability, th stands for throughput. For example, the curve labeled as (Tout=4, theory, th) means that the curve is the normalized throughput calculated by the model discussed above, where timeout value

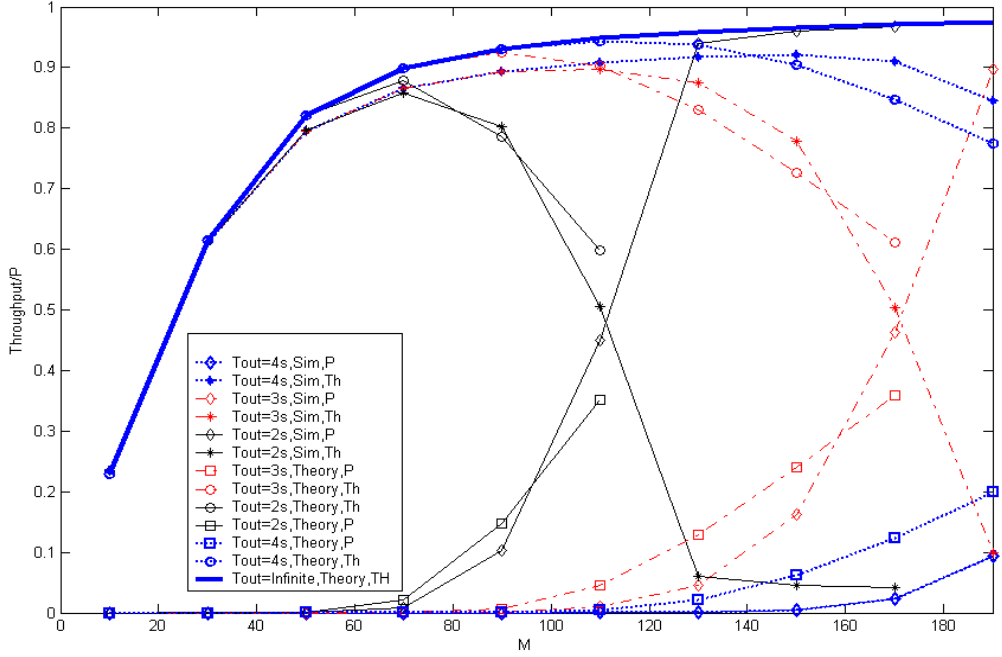

Figure 4 Analytical and Simulation Results.

is 4 seconds.

The curve labeled as (Tout=infinity, Theory, TH) represents the throughput of M2M transport protocol without timeout mechanism and was obtained by using the model proposed in this paper. As M increases, the throughput approaches to 1. However, when M is larger than 40, the increasing in throughput will be very small.

Now we consider the curve labeled as (Tout=2s,Sim,Th) in Figure 4. This curve represents normalized throughput on the access link when the timer value is 2 seconds. This curve is gotten by simulation with varying number of threads from 10 to 170. From this figure, the throughput is increased with M until 70. This can be easily understood because the more threads work in pipeline, the higher data rate can be achieved. But after M is larger than 70, the throughput begins to decrease with increasing M. As the data packet rate approaches the bandwidth of access link, timeout probability of data packets begins to increase, which can be confirmed by curve labeled as (Tout=2s,Sim,P). That means there are more timeout packets wasting the bandwidth of access link. So, for timeout value to be 2 seconds, the optimal M to realize the maximum throughput is 70. The curve labeled as (Tout=2s,Theory,th) represents the throughput calculated by analytical model under the same settings in simulation.  From Figure 4, it is seen that our model fits to the simulation results very well. When the timeout values are 3, 4 seconds, The M2M has similar behavior. The optimal M will be 70, 90, 110, respectively, when the timeout values are 2, 3, 4 seconds.

Ref. [1] describes that there is an optimal number of neighbors in PPLive and this optimal number of neighbors was claimed to be about 30 to 40 without explanation. We assume down/up bandwidth is 512kbps/256kbps, timeout value is 4 seconds, Tp is 600 ms respect. Using the modified model in the section 2.4, we found that the optimal M is 35. If each peer downloads data from each of other peers with one UDP thread, our model can give this phenomenon an explanation. However, since the rate of PPLive streaming is 400 kbps and the bandwidth of uplink of ADSL is only 256 kbps, we have to assume there are super nodes in a PPLive system to provide extra bandwidth.

## 4 Conclusion

In this paper, we argue that it is necessary to build a M2M protocol layer in P2P data distribution systems. For this purpose, we proposed a M2M transport protocol model and conducted related analysis. Numerical results were used to validate the effectiveness of this model. From this model, we found there exists an optimal download rate for given network settings and delay requirement.

**References**
[1] Yan Huang, Tom Z. J. Fu, Dah-Ming Chiu, John C. S. Lui, and Cheng Huang, Challenges, Design and Analysis of a Large-Scale P2P-VoD System, in *Proceedings of ACM SIGCOMM 2008*.
[2] Floyd, S., and Fall, K., Promoting the Use of End-to-End Congestion Control in the Internet, *IEEE/ACM Transactions on Networking*, August 1999.
[3] Xiaoyuan Yang, Minas Gjoka, Parminder Chhabra, Athina Markopoulou, Pablo Rodriguez, Kangaroo. Video Seeking in P2P Systems. *8th International workshop on Peer-to-peer systems*, April 22–24, 2009
[4] Fangming Liu, Bo Li, and Lili Zhong, How P2P Streaming Systems Scale Over Time Under a Flash Crowd? *8th International workshop on Peer-to-peer systems*, April 22–24, 2009
[5] http://www.dslreports.com/shownews/Using-The-New-UDP-BitTorrent-To-Dodge-ISP-Throttling-99514
[6] Min Zhang, Maurizio Dusi, Wolfgang John and Changjia Chen, Analysis of UDP Traffic Usage on Internet Backbone Links, *2009 Ninth Annual International Symposium on Applications and the Internet*.
[7] Mischa Schwartz, Telecommunication networks protocol, Modeling and analysis, Assison-Wesley publishing company, 1987.
[8] Ns Network Simulator (version 2.27). URL: http://www-mash.cs.berkeley.edu/ns/.
[9] http://nrg.cs.ucl.ac.uk/mptcp
[10] Chunxi Li, Changjia Chen. "Measurement Based PPStream Client Behavior Analysis". In Proc. of 2009 International Colloquium on Computing, Communication, Control, and Management, Aug 2009

[11] Vijay Vasudevan, Amar Phanishayee, etc, "Safe and Effective Fine-grained TCP Retransmissions for Datacenter Communication", Sigcomm 2009